\documentclass[12pt]{article}
\newcommand{\bra}{\langle}
\newcommand{\ket}{\rangle}
\newcommand{\R}{\mbox{\boldmath $ R $}}

\newcommand{\Z}{\mbox{\boldmath $ Z $}}

\def\dfrac#1#2{\displaystyle\frac{#1}{#2}}

\newcommand{\mapright}[2]
{\mathop{\hbox to 8mm{\rightarrowfill}}
\limits^{\scriptstyle #1}_{\scriptstyle #2}}

\makeatletter
\@addtoreset{equation}{section}
\makeatother

\begin{document} 
\begin{flushright}
hep-th/0105196 \\
KOBE-TH-01-03 \\
May 21, 2001 
\end{flushright}
\vspace*{6mm}
\begin{center}
{\Large \bf 
Spontaneous breaking of the rotational symmetry \vspace{2mm}\\
induced by monopoles in extra dimensions}
\vspace{10mm} \\
Seiho Matsumoto\footnote{e-mail: matsumoto@octopus.phys.sci.kobe-u.ac.jp},
Makoto Sakamoto\footnote{e-mail: sakamoto@phys.sci.kobe-u.ac.jp},
Shogo Tanimura\footnote{e-mail: tanimura@kues.kyoto-u.ac.jp}
\vspace*{10mm} \\
{\small \it
${}^1 $Graduate School of Science and Technology, Kobe University, 
Rokkodai, Nada, \\ Kobe 657-8501, Japan
\\
${}^2 $Department of Physics, Kobe University, 
Rokkodai, Nada, Kobe 657-8501, Japan
\\
${}^3 $Department of Engineering Physics and Mechanics, 
Kyoto University, \\ Kyoto 606-8501, Japan
}
\vspace{10mm}
\\
Abstract
\vspace{4mm}
\\
\begin{minipage}[t]{130mm}
\baselineskip 6mm 
We propose a field theoretical model 
that exhibits spontaneous breaking of the rotational symmetry.
The model has a two-dimensional sphere as extra dimensions of the space-time
and consists of a complex scalar field and a background gauge field.
The Dirac monopole,
which is invariant under the rotations of the sphere,
is taken as the background field.
We show that 
when the radius of the sphere is larger than a certain critical radius,
the vacuum expectation value of 
the scalar field develops vortices,
which pin down the rotational symmetry to lower symmetries.
We evaluate the critical radius 
and calculate configurations of the vortices by the lowest approximation.
The original model has a $ U(1) \times SU(2) $ symmetry and
it is broken to $ U(1) $, $ U(1) $, $ D_3 $ 
for each case of the monopole number $ q = 1/2, 1, 3/2 $,
respectively,
where $ D_3 $ is the symmetry group of a regular triangle.
Moreover, we show that the vortex configurations are stable against
higher corrections of the perturbative approximation.
\end{minipage}
\end{center}
\vspace{8mm}
\newpage
\baselineskip 6mm 

\section{Introduction}
Explorations of extra dimensions of the space-time
have a rich history and is recently calling new interests of the physicists
since, for example, existence of extra dimensions suggests a solution
to the hierarchy problem \cite{Randall}.
Moreover, the latest developments in a brane world scenairo \cite{brane}
are clarifying that Nambu-Goldstone bosons appear as a result
of spontaneous breaking of the translational symmetry in extra 
dimensions and that they play a role of a probe into the extra dimensions.
Recently, some of the authors \cite{Sakamoto}
constructed field theoretical models 
that exhibit spontaneous breaking of the translational symmetry
in extra one dimension.

One of our motivations to study models of translational symmetry breaking
is that they may function as a mechanism to break supersymmetry.
Since the algebra of supersymmetric charges generates
the algebra of translations,
translational symmetry breaking involves supersymmetry breaking.
Actually, some of the authors \cite{Takenaga} have constructed models
in which supersymmetry breaking is induced 
by spontaneous breaking of the translational symmetry.
However, the previous studies are restricted to 
the models which have only one extra dimension, that is, $ S^1 $.
To get a realistic model
it is desirable to make models with higher extra dimensions,
which have richer particle spectra.

In this Letter, 
we construct a model with two extra dimensions
that are represented by $ S^2 $.
This model consists of a complex scalar field and
has the rotational symmetry described by the group $ SU(2) $.
As a seed that causes symmetry breaking, 
we put a background monopole field in $ S^2 $.
We show that there is a critical radius of $ S^2 $:
When the radius of $ S^2 $ exceeds the critical radius,
the scalar field exhibits a vacuum expectation value
that is not uniform over $ S^2 $,
and therefore, the rotational symmetry is spontaneously broken.
In the following,
we will give the definition of our model and evaluate the critical radius.
We will examine structure of vacua of this model 
for cases of a few monopole numbers, 
and show that the $ SU(2) $ symmetry is broken to
$ U(1) $, $ U(1) $, $ D_3 $ for each case of the monopole number
$ q = 1/2, 1, 3/2 $, respectively.
Here $ D_3 $ denotes the symmetry group of a regular triangle.

\section{Model}
{}First, we define a model which exhibits
spontaneous breaking of the rotational symmetry.
Our model is defined in a space-time $ S^2 \times M^n $,
where $ S^2 $ is a two-dimensional sphere of a radius $ r $
and $ M^n $ is an $ n $-dimensional Minkowskian space.
The spherical coordinate of $ S^2 $ is denoted by $ (\theta,\phi) $,
and the Cartesian coordinate of $ M^n $ is denoted by 
$ (x^0,x^1, \cdots, x^{n-1} ) $.
The space $ M^n $ is equipped with
the metric $ g_{\mu \nu} = \mbox{diag} (+1,-1,\cdots,-1) $.
Our model consists of a complex scalar field $ f $ over $ S^2 \times M^n $
with a background gauge field $ A $ over $ S^2 $.
The gauge field $ A $ is fixed to be the Dirac-Wu-Yang monopole \cite{Wu},
which is defined as follows:
Two open sets of $ S^2 $,
$ U_+ = \{ (\theta,\phi) | \theta \ne \pi \} $ and
$ U_- = \{ (\theta,\phi) | \theta \ne 0   \} $,
cover $ S^2 = U_+ \cup U_- $.
The monopole field $ A $ is described by the pair of fields $ (A_+,A_-) $
which consists of the components
\begin{equation}
        A_{\pm}(\theta,\phi) = q ( \pm 1 - \cos \theta ) d \phi \quad 
		\mbox{in} \; U_{\pm},
        \label{gauge}
\end{equation}
respectively.
The magnetic field is given by
$ B = dA_+ = dA_- = q \sin \theta \, d \theta \wedge d \phi $.
Then the total magnetic flux is given by 
$ {\mit\Phi} = \int B = 4 \pi q $.
The scalar field $ f $ is described by a pair of fields $ (f_+, f_- ) $,
where $ f_\pm $ is a complex field over $ U_\pm $, respectively.
The pairs of the fields,
$ (A_+, f_+) $ and $ (A_-,f_-) $, are patched together 
by the gauge transformation 
\begin{equation}
        A_- = A_+ - 2q \, d \phi,
        \qquad
        f_- = e^{ - 2 i q \phi } f_+ 
        \label{patch}
\end{equation}
in $ U_+ \cap U_- $.
To make the field $ f_\pm $ single-valued,
$ 2q $ must be an integer.
The covariant derivative of the field $ f $ is defined as 
$ Df_\pm = df_\pm - iA_\pm f_\pm $.
The action of our model is given by
\begin{eqnarray}
        S_A [f] &=&
        \int d^n x\, d \theta d \phi \, r^2 \sin \theta   
        \Bigg\{
                - \frac{1}{r^2}
                \Bigg|
                        \frac{\partial f_{\pm}}{\partial \theta}
                \Bigg|^2
                -
                \frac{1}{r^2 \sin^2 \theta}
                \Bigg|
                        \frac{\partial f_{\pm}}{\partial \phi}
                        - i q
                        ( \pm 1 - \cos \theta ) f_{\pm}
                \Bigg|^2
        \nonumber \\ && \qquad \qquad \qquad \qquad
                +
                g^{\mu \nu}
                \frac{\partial f_{\pm}^*}{\partial x^\mu}
                \frac{\partial f_{\pm}  }{\partial x^\nu}
                + \mu^2 f_{\pm}^* \! f_{\pm}
				- \lambda (f_{\pm}^* \! f_{\pm})^2
        \Bigg\}
        \label{action}
\end{eqnarray}
with the real parameters $ \mu^2, \lambda > 0 $.

This model has a global symmetry $ U(1) \times SU(2) $.
The $ U(1) $ symmetry is defined as a family of the transformations
\begin{equation}
        ( f_+, f_- ) \to e^{it} ( f_+, f_- )
        \label{U(1)}
\end{equation}
with $ t \in \R $.
On the other hand,
the elements of $ SU(2) $ act on $ S^2 $ as rotation transformations.
Of course, the explicit form of the background gauge field 
$ A_\pm = q ( \pm 1 - \cos \theta ) d \phi $
is not invariant under arbitrary rotations.
Actually,
to leave the background field 
$ A_{\pm} $ invariant,
a rotation transformation is needed
to be combined with a gauge transformation.
Such a gauge transformation can be found by using the Wigner rotation,
which is a well-known technique 
in the theory of induced representations \cite{Mackey}.
To describe the $ SU(2) $ symmetry 
we need to provide some notation as follows:
The usual spherical coordinate $ (\theta,\phi) $ 
is assigned to a point $ p \in S^2 $.
We define two maps $ s_\pm : U_\pm \to SU(2) $ as
\begin{equation}
        s_\pm (p) :=
        e^{-i \sigma_3 \phi  /2}
        e^{-i \sigma_2 \theta/2}
        e^{\pm i \sigma_3 \phi  /2}.
        \label{sections}
\end{equation}
Let indices $ \alpha $ and $ \beta $ denote either $ + $ or $ - $.
Suppose that a point 
$ p \in U_\alpha $ is transformed to
$ g^{-1} p \in U_\beta $ by an element $ g \in SU(2) $.
Then the Wigner rotation is defined as
\begin{equation}
        W_{\alpha \beta} (g;p) :=
        s_\alpha(p)^{-1} \cdot g \cdot s_\beta(g^{-1} p).
        \label{Wigner}
\end{equation}
It is easily verified that the value of the Wigner rotation has a form 
\begin{equation}
        W_{\alpha \beta} (g;p) =
        e^{-i \sigma_3 \omega /2}
        \label{Wigner in U(1)}
\end{equation}
with a real number $ \omega $,
which defines a function $ \omega_{\alpha \beta} (g;p) $.
Using it we define the transformation of the fields by $ g \in SU(2) $ as
\begin{eqnarray}
        A_\alpha(p) 
        & \to & A'_\alpha(p) 
        = A_\beta(g^{-1}p) + q \, d \omega_{\alpha \beta} (g;p),
        \label{SU(2)A} \\
        f_\alpha(p) & \to & f'_\alpha(p) 
        = e^{iq \omega_{\alpha \beta} (g; p)} f_\beta(g^{-1} p).
        \label{SU(2)f}
\end{eqnarray}
Under this transformation
the monopole background field remains invariant as $ A'_\pm = A_\pm $,
and the action (\ref{action}) is also left invariant.

Although calculation of the concrete value of 
the Wigner rotation (\ref{Wigner}) is cumbersome,
to give a definite example we calculate it for 
$ g = e^{-i \sigma_3 \gamma/2} $,
which is a rotation around the $ z $-axis.
The point $ p = (\theta, \phi) \in U_\pm $ is then transformed to
$ g^{-1} p = (\theta, \phi - \gamma) \in U_\pm $, 
and from the definitions (\ref{sections}) and (\ref{Wigner}) we get
\begin{equation}
        W_{\pm \pm} ( e^{-i \sigma_3 \gamma/2} ;p) 
        = 
        e^{\mp i \sigma_3 \gamma /2},
        \label{Wigner for z-rotation}
\end{equation}
which implies $ \omega_{\pm \pm}(g; p) = \pm \gamma = $ constant
in (\ref{Wigner in U(1)}).
The transformations (\ref{SU(2)A}) and (\ref{SU(2)f}) now become
\begin{eqnarray}
        A_\pm (\theta,\phi) 
& \to & 
        A'_\pm (\theta,\phi) 
        = A_\pm (\theta,\phi - \gamma) \pm q \, d \gamma
        = A_\pm (\theta,\phi), \qquad
        \label{A=A} \\
        f_\pm (\theta,\phi) 
& \to & 
        f'_\pm (\theta,\phi) 
        = e^{\pm iq \gamma} f_\pm (\theta,\phi - \gamma).
        \label{f to f'}
\end{eqnarray}
Thus the invariance of $ A_\pm $ is checked and 
the patching condition (\ref{patch})
is also satisfied by $ (f'_+, f'_- ) $.

{}For later use, let us calculate the Wigner rotation (\ref{Wigner}) 
for the rotation around the $ x $-axis with the angle $ \pi $,
which is given by
$ g = e^{-i \sigma_1 \pi/2} = -i \sigma_1 $.
The point $ p = (\theta, \phi) \in U_\pm $ is then transformed to
$ g^{-1} p = (\pi-\theta, -\phi) \in U_\mp $.
Then the corresponding Wigner rotation is
\begin{equation}
        W_{\pm \mp} ( e^{-i \sigma_1 \pi/2} ;p) 
        =
        -i \sigma_3 
        =
        e^{ -i \sigma_3 \pi/2 },
        \label{Wigner for x-rotation}
\end{equation}
and therefore we have $ \omega_{\pm \mp}(g; p) = \pi $ 
in the place of (\ref{Wigner in U(1)}).
The transformations (\ref{SU(2)A}) and (\ref{SU(2)f}) now become
\begin{eqnarray}
        A_\pm (\theta,\phi) 
& \to & 
        A'_\pm (\theta,\phi) 
        = A_\mp (\pi-\theta, -\phi) 
        = A_\pm (\theta,\phi), \qquad
        \label{A=A x-axis} 
        \\
        f_\pm (\theta,\phi) 
& \to & 
        f'_\pm (\theta,\phi) 
        = e^{iq \pi} f_\mp (\pi-\theta, -\phi).
        \label{f to f' x-axis}
\end{eqnarray}
Thus the gauge field $ A_\pm $ remains invariant again.

\section{Rotational symmetry breaking}
Assume that the monopole number $ q $ is not zero
and that the component fields $ (f_+,f_-) $ are continuous functions.
Then it is easily proved that
if the field $ f $ is rotationally invariant, 
it must vanish identically as $ f \equiv 0 $ over $ S^2 $.
The contraposition says that
if $ f $ is not identically zero, 
$ f $ cannot be rotationally invariant over $ S^2 $.
Actually, even if $ f $ takes nonzero values in some region,
the value of $ f (\theta,\phi) $ 
must vanish generally at $ 2|q| $ points in $ S^2 $.
Hence the zero points of $ f $ pin down the rotational symmetry.
A zero point of $ f $ is called a vortex.

Here we describe only the outline of the proof of the above theorem.
The statement that $ f $ is rotationally invariant means that
both $ f_\pm $ remain invariant under the transformations (\ref{SU(2)f}) 
by $ SU(2)$.
If $ f $ is rotationally invariant and takes nonzero value 
at some point in $ S^2 $,
$ f_\pm $ should be nonvanishing everywhere
because the group $ SU(2) $ acts on $ S^2 $ transitively.
Suppose that $ f_+ $ does not vanish in the upper hemisphere,
$ 0 \le \theta \le \pi/2 $.
When the coordinate $ \phi $ runs over the range $ 0 \le \phi \le 2 \pi $ 
with $ \theta $ fixed at $ \theta = \pi/2 $,
the value of $ f_- = e^{-2iq \phi} f_+ $
runs around the zero in the complex plane $ 2 q $ times.
Thus the value of the continuous function $ f_- $ becomes zero as its value
at least $ 2|q| $ times in the lower hemisphere, $ \pi/2 \le \theta \le \pi $.
A sketch of the proof is then over.

The above theorem tells that 
if the scalar field exhibits a nonzero vacuum expectation value $ \bra f \ket $,
the rotational symmetry is necessarily broken.
Now we would like to examine a condition for rotational symmetry breaking.
In this paper we analyze the model only 
at the classical level.
In other words, we will seek for a vacuum field configuration
which minimizes the classical energy.
Moreover, since the translational symmetry in $ M^n $ is kept unbroken,
what we need to find is the vacuum field $ f (\theta, \phi) $ 
that minimizes the classical energy functional
\begin{equation}
        E 
= 
        \int d \theta d \phi \, r^2 \sin \theta
        \Bigg\{
                \frac{1}{r^2}
                \Bigg|
                        \frac{\partial f_\pm}{\partial \theta}
                \Bigg|^2
                +
                \frac{1}{r^2 \sin^2 \theta}
                \Bigg|
                        \frac{\partial f_\pm}{\partial \phi}
                        - iq
                        ( \pm 1 - \cos \theta ) f_\pm
                \Bigg|^2
                - \mu^2 f_{\pm}^* \! f_{\pm}
				+ \lambda (f_{\pm}^* \! f_{\pm})^2
        \Bigg\}.
        \label{energy}
\end{equation}
Variation of the gradient energy with respect to $ f_\pm^* $ gives
the Laplacian coupled with the monopole,
\begin{equation}
        - \Delta_q f_\pm
:=
        - \left[
                \frac{1}{\sin \theta}
                \frac{\partial}{\partial \theta}
                \bigg(
                        \sin \theta
                        \frac{\partial}{\partial \theta}
                \bigg)
                +
                \frac{1}{\sin^2 \theta}
                \bigg(
                        \frac{\partial}{\partial \phi}
                        -iq ( \pm 1 - \cos \theta )
                \bigg)^2
        \right] f_\pm.
        \label{Laplacian}
\end{equation}
The eigenvalue problem of the monopole Laplacian has been solved by 
Wu and Yang \cite{Wu};
its eigenfunctions are expressed in terms of
the matrix elements of unitary representations
of $ SU(2) $ as 
\begin{equation}
        f_\pm (\theta,\phi)
        =
        D^j_{mq} (\theta, \phi, \mp \phi)
        =
        \bra j, m | 
        e^{-i J_3 \phi} \, e^{-i J_2 \theta} \, e^{\pm i J_3 \phi} 
        | j,q \ket
        =
        e^{-i (m \mp q) \phi} \, d^j_{mq} (\theta)
        \label{eigenfunction}
\end{equation}
which has the eigenvalue
\begin{equation}
        \epsilon_j =  j(j+1) - q^2,
        \qquad
        ( j = |q|, |q|+1, |q|+2, \cdots ).
        \label{eigenvalue}
\end{equation}
Each eigenvalue is degenerate with respect to the index $ m $,
which has a range $ m = -j, -j+1, \cdots, j-1, j $.
Since the lowest eigenvalue of the Laplacian is $ \epsilon_{|q|} = |q| $,
the lower bound of the energy is given by
\begin{equation}
        E 
\ge
        ( |q| - \mu^2 r^2 )
        \int d \theta d \phi \sin \theta | f_\pm |^2
        + \lambda r^2
        \int d \theta d \phi \sin \theta | f_\pm |^4.
        \label{energy bound}
\end{equation}
If $ |q| - \mu^2 r^2 > 0 $, the inequality
(\ref{energy bound}) is positive definite.
Then the minimum of $ E $ is realized only by the trivial vacuum 
$ f \equiv 0 $.
On the contrary,
if $ |q| - \mu^2 r^2 < 0 $, 
it is possible to find a field $ f $ which has a negative energy.
{}For example, take the eigenfunction (\ref{eigenfunction}) 
of the lowest eigenvalue,
$ f_\pm (\theta,\phi) = c \, e^{-i(m \mp q) \phi} \, d^{|q|}_{mq} (\theta) $,
where $ c $ is a complex number.
The energy of this field configuration
can be written into the form
\begin{equation}
        E 
= 
        - a_2 |c|^2 + a_4 |c|^4 
        \label{negative energy}
\end{equation}
with coefficients $ a_2 , a_4 > 0 $.
Thus, for $ 0 < |c|^2 < a_2/a_4 $, this field realizes $ E < 0 $.
Therefore, the minimum value of $ E $ must be negative
and it is realized by a nontrivial vacuum $ f \ne 0 $
for $ |q| - \mu^2 r^2 < 0 $.
Thus we conclude that 
the rotational symmetry is spontaneously broken 
when the radius $ r $ of $ S^2 $ is larger than the critical radius $ r_q $,
i.e.
\begin{equation}
        r > r_q := \frac{\sqrt{|q|}}{\mu}.
        \label{critical radius}
\end{equation}
We may rephrase this condition 
by borrowing the language of superconductivity.
When the scalar field $ f $ develops a nonzero vacuum expectation value
in the Lagrangian (\ref{action}), 
the Higgs boson with the mass $ M_\sigma^2 = 2 \mu^2 $ would appear.
The coherence length $ \xi $ is defined as $ \xi = M_\sigma^{-1} $,
which characterizes the radius of the core of a vortex.
In terms of $\xi$, 
the condition for rotational symmetry breaking can be expressed as
\begin{equation}
        r > \sqrt{2|q|} \, \xi.
        \label{coherence length}
\end{equation}

\section{Vacuum configurations}
We shall now examine the concrete vacuum configuration of the scalar field.
We will obtain approximate solutions of the scalar field
by a variational method
for three cases with the monopole number $ q = 1/2, 1, 3/2 $.
Let us explain briefly our method of calculation.
A generic field which satisfies the patching condition (\ref{patch})
can be expanded in a series of the eigenfunctions (\ref{eigenfunction}) as
\begin{equation}
        f_{\pm}^{q} (\theta, \phi)
        =
        \sum_{j=|q|}^\infty
        \sum_{m = -j}^j
        c^j_m \, D^{j}_{m,q} (\theta, \phi, \mp \phi).
        \label{expansion}
\end{equation}
Here we use the lowest approximation for it; 
we restrict the above series to the leading terms
\begin{equation}
        f_{\pm}^{q} (\theta, \phi)
        =
        \sum_{m = -|q|}^{|q|}
        c_m \, D^{|q|}_{m,q} (\theta, \phi, \mp \phi)
        \label{lowest expansion}
\end{equation}
and substitute it into the energy functional (\ref{energy}).
Then the vacuum configuration is determined by
the coefficients $ \{ c_m \} $ that minimize the energy.
In the next section this lowest approximation will be justified.

When the monopole has the smallest charge $ q = 1/2 $,
in the lowest expansion (\ref{lowest expansion})
we put the coefficients as
$ c_{1/2}  = - v e^{ i \gamma /2} \sin (\beta/2) $ and
$ c_{-1/2} =   v e^{-i \gamma /2} \cos (\beta/2) $
with the real parameters $ (v, \beta, \gamma) $.
Then we have
\begin{equation}
        f_{\pm}^{1/2} (\theta, \phi)
        = 
        v \biggl[
                -e^{-i (\phi-\gamma)/2}
                \sin (\beta/2) \cos ( \theta/2 ) 
                +
                e^{i (\phi-\gamma)/2} 
                \cos (\beta/2) \sin ( \theta/2 ) 
        \biggr]
        e^{\pm i \phi/2}.
        \label{ground state for q=1/2}
\end{equation}
Thus it can easily be seen that
the value of $ f^{1/2} $ vanishes
at the point $ (\theta, \phi) = (\beta, \gamma) $.
Since the position of the zero point of $f^{1/2}$ can be moved
to the north pole of $S^2$ by using an $SU(2)$ rotation,
we can set $(\beta, \gamma) = (0, 0)$ without loss of generality.
Then
the field becomes simply
\begin{equation}
        f_{\pm}^{1/2} (\theta, \phi)
        = 
        v \, e^{i (1 \pm 1) \phi /2} \sin ( \theta/2 ),
        \label{ground for 1/2}
\end{equation}
and thus a single vortex is located at the north pole of $ S^2 $.
The energy (\ref{energy}) is then calculated as
\begin{equation}
        E 
        =
        - 2 \pi \left( \mu^2 r^2 - \frac{1}{2} \right) v^2
        + 
        \frac{2}{3} \cdot 2 \pi \lambda r^2 v^4.
        \label{variational energy for q=1/2}
\end{equation}
Hence the energy is minimized when
\begin{equation}
    v^2 = 
	\left\{
		\begin{array}{ll}
		    0 
		    & \quad \mbox{for} \quad r \leq (\sqrt{2} \, \mu)^{-1}, 
		    \\
		    \dfrac{3}{4\lambda r^2} 
		    \Big( \mu^2 r^2 -\frac{1}{2} \Big) 
		    & \quad \mbox{for} \quad r > (\sqrt{2} \, \mu)^{-1}.
		\end{array} 
	\right.
        \label{VEV for q=1/2}
\end{equation}
This result is to be compared with the vacuum expectation value
\begin{equation}
        \bra f^0 \ket^2 
        = \frac{1}{2 \lambda} \mu^2
        \label{VEV for q=0}
\end{equation}
for the case of $ q=0 $.

It can also be verified that
the vacuum field configuration (\ref{ground for 1/2}) is invariant under
the combined transformation 
of the phase (\ref{U(1)}) 
with the rotation (\ref{f to f'}) of the angle $ \gamma = 2t $,
\begin{equation}
        ( f_+^{1/2}, f_-^{1/2} ) (\theta,\phi)
        \; \to \;
        e^{it} ( e^{it} f_+^{1/2}, e^{-it} f_-^{1/2} )
        (\theta,\phi - 2t).
        \label{remaining U(1)}
\end{equation}
Therefore, we conclude that when the monopole number is $ q = 1/2 $
and $r>r_{1/2}$,
the symmetry $ U(1) \times SU(2) $ is spontaneously broken to the subgroup 
$ U(1) $ which is defined by (\ref{remaining U(1)}).
Then three massless Nambu-Goldstone bosons 
appear due to the symmetry breaking.
However, one of the three would be absorbed into the gauge boson
via the Higgs mechanism 
if the gauge field has its own dynamical degrees of freedom.

Next let us consider the case of $ q=1 $.
Then a generic scalar field $ f $ has two vortices on $ S^2 $.
The lowest expansion (\ref{lowest expansion}) now becomes
\begin{equation}
        f_{\pm}^1 (\theta, \phi)
= 
        \frac{1}{2}
        \left[
                c_{+1} \, e^{-i \phi} ( 1 + \cos \theta ) 
                +
                c_{0}  \, \sqrt{2} \sin \theta
                +
                c_{-1} \, e^{i \phi} ( 1 - \cos \theta ) 
        \right] e^{\pm i \phi} .
        \label{ground state for q=1}
\end{equation}
By using the $ U(1) \times SU(2) $ symmetry,
which is defined in (\ref{U(1)}) and (\ref{SU(2)f}),
it is always possible to
bring the coefficients into the form
\begin{equation}
        ( c_{+1}, c_0, c_{-1}  )
=
	( 
		\frac{v}{\sqrt{2}}\sin\alpha, 
		v \cos \alpha,
		\frac{v}{\sqrt{2}}\sin\alpha
	)
        \label{c vector}
\end{equation}
with the real parameters $ v $ and $ \alpha\ (0\leq \alpha \leq \pi/4) $.
Then the field (\ref{ground state for q=1}) becomes
\begin{equation}
        f_{\pm}^1 (\theta, \phi)
= 
        \frac{1}{\sqrt{2}} \, v
        \Bigl[
                  ( \cos \alpha \sin \theta + \sin \alpha \cos \phi )
                - i \sin \alpha \cos \theta \sin \phi 
        \Bigr] e^{\pm i \phi}.
        \label{ground state for q=1 by alpha}
\end{equation}
Thus two zero points of $f^1$ are located at
$ (\theta, \phi) = (\sin^{-1}(\tan\alpha), \pi) $.
So, the relative displacement of two vortices is changed 
by the parameter $ \alpha $.

When the trial function (\ref{ground state for q=1 by alpha}) is substituted,
the total energy (\ref{energy}) is evaluated as
\begin{equation}
        E 
        =
        \frac{4 \pi}{3} ( 1 - \mu^2 r^2 ) v^2
        + \frac{8 \pi}{15} 
        \left\{
                1 + \frac{1}{4} ( 1 - \cos 4 \alpha )
        \right\}
        \lambda r^2 v^4.
        \label{total energy of f_1}
\end{equation}
By changing $ \alpha $,
the minima of the potential is realized when
$ \cos 4 \alpha = 1 $, i.e. 
$ \alpha = 0 $.
Then two vortices are located at opposite two points on $ S^2 $.
On the other hand, the maxima of the potential is realized when
$ \cos 4 \alpha = -1 $, i.e. 
$ \alpha = \pi/4$.
Then two vortices coincide.
Thus we observe that the vortices repel each other.
The minimum of the energy is realized by $ \alpha = 0 $ with
\begin{equation}
        v^2 
        = \frac{5}{4 \lambda} \left( 1 - \frac{1}{\mu^2 r^2} \right) \mu^2
        \quad \mbox{for} \quad
	r > \frac{1}{\mu}.
        \label{VEV of q=1}
\end{equation}
The vacuum configuration (\ref{ground state for q=1 by alpha}) 
then becomes
\begin{equation}
        f_{\pm}^1 ( \theta, \phi)
        =
        \frac{1}{\sqrt{2}} v \sin \theta \, e^{\pm i \phi},
        \label{opposite vortex}
\end{equation}
and has two vortices at the north and the south poles of $ S^2 $.
This configuration (\ref{opposite vortex})
is invariant under the rotations around the $ z $-axis (\ref{f to f'}).
Thus, we conclude that when the monopole number is $ q = 1 $,
the symmetry $ U(1) \times SU(2) $ is spontaneously broken to $ U(1) $
for $ r > \mu^{-1} $.
At this time also
three Nambu-Goldstone bosons appear,
although one of the three would disappear via the Higgs mechanism.

{}Finally, let us examine the case of $ q=3/2 $.
Then the number of vortices is three.
By similar but tedious calculation
we can determine the coefficients in the expansion (\ref{lowest expansion}).
Here we show only the result: 
The minimum energy configuration is given,
up to the $ U(1) \times SU(2) $ symmetry, by
\begin{eqnarray}
        f_{\pm}^{3/2} 
        (\theta,\phi)
& = &
        v 
        \Big[
                  e^{-i \phi /2} \cos ( \theta/2 ) 
                - e^{ i \phi /2} \sin ( \theta/2 ) 
        \Big]
        \nonumber \\ && 
        \times \Big[
                  e^{-i (\phi - 2 \pi/3) /2} \cos ( \theta/2 ) 
                - e^{ i (\phi - 2 \pi/3) /2} \sin ( \theta/2 ) 
        \Big]
        \nonumber \\ && 
        \times \Big[
                  e^{-i (\phi - 4 \pi/3) /2} \cos ( \theta/2 ) 
                - e^{ i (\phi - 4 \pi/3) /2} \sin ( \theta/2 ) 
        \Big]
        e^{\pm 3 i \phi /2 } 
        \label{vacuum of q=3/2}
\end{eqnarray}
with
\begin{equation}
        v^2 
        =
        \frac{35}{44 \lambda}
        \left( 1 - \frac{3}{2 \mu^2 r^2} \right)
        \mu^2 
        \quad \mbox{for} \quad
        r > \sqrt{\frac{3}{2}} \, \frac{1}{\mu}.
        \label{v of q=3/2}
\end{equation}
We can read off the location of zero points from (\ref{vacuum of q=3/2});
they are located at $ \phi = 0, 2\pi/3, 4\pi/3 $
on $ \theta = \pi/2 $.
Namely, the vortices are located at the vertices
of the largest equilateral triangle on $ S^2 $.

Now we shall describe the symmetry of the vacuum.
By the combination of the rotation around the $ z $-axis (\ref{f to f'})
of the angle $ \gamma = 2 \pi/3 $
with the $ U(1) $ transformation (\ref{U(1)}) 
of the phase $ t = \pi $,
the scalar field is transformed accordingly as
\begin{equation}
        ( f_+^{3/2}, f_-^{3/2} ) (\theta,\phi)
        \mapright{R}{} 
        ( f_+^{3/2}, f_-^{3/2} ) (\theta,\phi - 2 \pi/3).
        \label{R}
\end{equation}
This transformation is denoted by $ R $ 
and it generates the cyclic group $ \Z_3 $.
Actually, the configuration (\ref{vacuum of q=3/2}) remains invariant under
the operation of $ R $.
Moreover, 
under a combination of 
the $ \pi $-rotation around the $ x $-axis (\ref{f to f' x-axis})
with the $ U(1) $ transformation (\ref{U(1)}) 
of the phase $ t = - \pi/2 $,
the scalar field changes as
\begin{equation}
        ( f_+^{3/2}, f_-^{3/2} ) (\theta,\phi)
        \mapright{T}{} 
        e^{i \pi} 
        ( f_-^{3/2}, f_+^{3/2} ) (\pi-\theta, -\phi),
        \label{T}
\end{equation}
and 
this transformation $ T $ generates another cyclic group $ \Z_2 $.
It can easily be verified that the configuration
(\ref{vacuum of q=3/2}) remains invariant also under the operation of $ T $.
Notice that two operations $ R $ and $ T $ do not commute each other;
they generate a nonabelian group $ D_3 $, 
which is called the dihedral group of the order three,
i.e. the symmetry group of a regular triangle.
We thus conclude that
when the monopole number is $ q = 3/2 $,
the symmetry $ U(1) \times SU(2) $ is spontaneously broken to 
the discrete nonabelian group $ D_3 $ 
for $ r > \sqrt{3/2} \, \mu^{-1} $.
Therefore,
the number of the Nambu-Goldstone bosons associated with the symmetry breaking
is four,
although one of the four would be absorbed into the gauge boson
via the Higgs mechanism.

\section{Stability of the vacuum}
In the previous section
we solved the problem to minimize the energy functional (\ref{energy})
by the variational method within the restricted function space
(\ref{lowest expansion}),
which is the lowest eigenspace of the monopole Laplacian (\ref{Laplacian}).
Here we would like to clarify validity of our analysis.

The first point to be clarified is 
that the critical radius (\ref{critical radius}) is exact
in the context of the classical field theory.
We did not recourse any approximation 
in the argument to decide the critical radius.

The second point to be noticed is 
accuracy of the concrete forms of the vacuum configurations, like
(\ref{ground for 1/2}),
(\ref{opposite vortex}),
(\ref{vacuum of q=3/2}).
They are approximately calculated by the variational method
with restricting
the full function space (\ref{expansion})
to the lowest eigenspace (\ref{lowest expansion})
of the monopole Laplacian.
This restriction is a good approximation
if the lowest eigenvalue $ \epsilon_{|q|} $ 
makes 
the quadratic term in (\ref{energy}) negative 
but the next eigenvalue $ \epsilon_{|q|+1} $ 
gives positive contribution to it.
More explicitly, 
referring to the eigenvalues (\ref{eigenvalue}),
the necessary condition for the validity is
\begin{equation}
        \epsilon_{|q|} - \mu^2 r^2 
        < 0 <
        \epsilon_{|q|+1} - \mu^2 r^2, 
        \quad \mbox{or} \quad
        \frac{\sqrt{|q|}}{\mu} 
        < r < 
        \frac{\sqrt{3|q| + 2}}{\mu}.
        \label{valid range}
\end{equation}
In other words,
the restricted variational method 
is a good approximation
when the radius of $ S^2 $ is larger than just the critical radius 
but not too large.

The third point to be examined is the stability of the vacuum 
against perturbation by higher eigenvalue functions.
The approximation is improved by including higher-order terms in the expansion
(\ref{expansion}).
Now a question arises:
If we include some of higher terms or all the terms in the expansion
(\ref{expansion}),
does the better approximated or the precise vacuum have the same symmetry 
as the lowest-approximated vacuum has?

The answer is affirmative:
When higher-order terms are included in the trial function
(\ref{expansion}),
but if the radius of $ S^2 $ is in the range (\ref{valid range}),
the vacuum calculated by the higher expansion has the same symmetry
as the vacuum calculated by the lowest approximation has.

The above statement is proved as follows:
{}First,
notice that the space of the scalar fields $ f = (f_+,f_-) $
provides a unitary representation of $ U(1) \times SU(2) $
by the transformations (\ref{U(1)}) and (\ref{SU(2)f}).
Let $ f^{(0)} $ denote the solution by the lowest approximation like
(\ref{ground for 1/2}),
(\ref{opposite vortex}),
(\ref{vacuum of q=3/2}).
Assume that $ f $ is a solution by the higher-order approximation
and define $ f^{(1)} $ as a correction to $ f^{(0)} $, i.e.
\begin{equation}
        f = f^{(0)} + f^{(1)}.
        \label{difference}
\end{equation}
Let $ H $ be a subgroup of $ U(1) \times SU(2) $
that preserves $ f^{(0)} $ invariant.
Then we decompose
$ f^{(1)} $
into a component
$ ( f^{(1)} )_\parallel $ that is in the identity representation of $ H $,
and its orthogonal complement
$ ( f^{(1)} )_\perp $
as 
\begin{equation}
        f^{(1)} =
        ( f^{(1)} )_\parallel + ( f^{(1)} )_\perp.
        \label{decompose}
\end{equation}
When (\ref{difference}) is substituted,
the energy functional (\ref{energy}) is symbolically written as
\begin{equation}
        E [ f^{(0)} + f^{(1)} ] 
        =
        E [ f^{(0)} ] 
        + \frac{\delta E}{\delta f} [ f^{(0)} ] \cdot f^{(1)}
        + \frac{1}{2} \frac{\delta^2 E}{\delta f^2} [ f^{(0)} ] 
        \cdot (f^{(1)})^2
        + \cdots.
        \label{expand E}
\end{equation}
The second term of the RHS is
\begin{eqnarray}
        \frac{\delta E}{\delta f} [ f^{(0)} ] \cdot f^{(1)}
        & = &
        \int d \theta d \phi \sin \theta
        \Big\{
                f^{(0)*} 
                ( - \Delta_q - \mu^2 r^2 
                + 2 \lambda r^2 | f^{(0)} |^2 ) 
                f^{(1)}
        \nonumber \\ && \qquad \qquad \quad
                +
                f^{(1)*} 
                ( - \Delta_q - \mu^2 r^2 
                + 2 \lambda r^2 | f^{(0)} |^2 ) 
                f^{(0)}
        \Big\}.
        \label{1st order}
\end{eqnarray}
Since $ f^{(0)} $ is invariant under the actions of $ H $,
$ ( - \Delta_q - \mu^2 r^2 + 2 \lambda r^2 |f^{(0)}|^2 ) f^{(0)} $ is also
invariant.
Therefore, the term linear in the orthogonal component $ (f^{(1)})_\perp $
vanishes as
\begin{equation}
        \frac{\delta E}{\delta f} [ f^{(0)} ] \cdot (f^{(1)})_\perp
        = 0.
        \label{orth}
\end{equation}
Moreover, the third term of the RHS of (\ref{expand E}) is
\begin{eqnarray}
        \frac{1}{2} \frac{\delta^2 E}{\delta f^2} [ f^{(0)} ] 
        \cdot ( f^{(1)} )^2
& = &
        \int d \theta d \phi \sin \theta 
        \bigg[
                f^{(1)*} 
                ( - \Delta_q - \mu^2 r^2 ) 
                f^{(1)}
                \nonumber \\ &&
                + \lambda r^2
                \Big\{
                          2 | f^{(0)} |^2 | f^{(1)} |^2
                        + \big( f^{(0)}  f^{(1)*} + f^{(0)*} f^{(1)} \big)^2
                \Big\}
        \bigg].
        \label{2nd order}
\end{eqnarray}
If the radius of $ S^2 $ is in the range (\ref{valid range}),
the first term in the last line is positive definite for $ ( f^{(1)} )_\perp $
since it is orthogonal to the space of the lowest-eigenvalue functions.
It is clear that the second term is also positive definite.
Thus we conclude that
the quadratic form (\ref{2nd order}) is positive for any $ ( f^{(1)} )_\perp $,
and therefore, 
the lowest-approximated vacuum $ f^{(0)} $ is stable 
against symmetry-breaking perturbation by $ ( f^{(1)} )_\perp $.
Hence in the vacuum $ f $ 
the orthogonal component vanishes,
i.e. $ ( f^{(1)} )_\perp = 0 $.
This implies that
the perturbed vacuum $ f = f^{(0)} + ( f^{(1)} )_\parallel $
remains 
invariant under the actions of $ H $.
The proof is over.

\section{Remarks}
We would like to put a remark about a relation of our argument with
the Coleman theorem \cite{Coleman},
which forbids spontaneous breaking of continuous symmetries
in two dimensions.
Our model is built in higher dimensions than two;
the space-time we concern is the direct product $ M^n \times S^2 $
of the Minkowski space $ M^n $ with the extra $ S^2 $.
Thus the Coleman theorem is not applicable to our model.

{}Finally, we shall briefly mention directions for further development
of our work.
Models with higher-dimensional manifolds than $ S^2 $ involve
higher symmetries and richer matter contents,
and such a generalization must be advantageous 
to construction of more realistic models.
In particular,
background gauge fields on higher-dimensional homogeneous spaces 
\cite{Tanimura} provide a relevant basis for further development.
It is also strongly desirable to build a supersymmetric model, 
in which translational or rotational symmetry breaking induces
supersymmetry breaking.
It remains as an important question to seek for a dynamical origin
of the background gauge field that triggers symmetry breaking.
We expect that the Hosotani mechanism \cite{Hosotani} or a similar mechanism
functions as the origin of the background field,
although it is not yet unveiled.

\section*{Acknowledgments}
The authors wish to thank 
H. Hatanaka, 
M. Hayakawa, 
Y. Hosotani, 
S. Iso, 
T. Kugo, 
C.S. Lim, 
Y. Nagatani, 
M. Nakahara, 
H. Nakano, 
K. Ohnishi, 
H. Otsu, 
N. Sakai, 
H. So,
M. Tachibana, 
K. Takenaga,
I. Tsutsui
and
K. Yamawaki
for valuable comments and criticisms.
This work was supported by Grant-in-Aids for Scientific Research 
({\#}12640275 for M.S. and {\#}12047216 for S.T.)
from
Ministry of Education, Culture, Sports, Science and Technology of Japan.

\baselineskip 5mm 

\end{document}